\documentclass[twocolumn,prl,showpacs,floatfix]{revtex4}

\usepackage{graphicx}
\usepackage{epsfig}
\usepackage{rotating}
\usepackage{amsmath}
\usepackage{amsfonts}
\usepackage{amssymb}
\usepackage{enumerate}
\usepackage{longtable}
\usepackage{dcolumn}
\usepackage{bm}

\begin{document}

\title{Spin-waves in the $J_{1a}-J_{1b}-J_{2}$ orthorombic square-lattice 
Heisenberg models: Application to the iron pnictide materials}

\author{R. Applegate}
\affiliation{University of California Davis, CA 95616, USA}

\author{J. Oitmaa }
\affiliation{School of Physics, The University of New South Wales,
Sydney 2052, Australia}

\author{R. R. P. Singh}
\affiliation{University of California Davis, CA 95616, USA}

\date{\rm\today}

\begin{abstract}
Motivated by the observation of 
spatially anisotropic exchange constants
in the iron pnictide materials, we study the spin-wave spectra of the
$J_{1a}-J_{1b}-J_{2}$ Heisenberg models on a square-lattice with
nearest neighbor exchange $J_{1a}$ along x and $J_{1b}$ along y axis and
a second neighbor exchange $J_2$. We focus on the regime, where the
spins order at ($\pi,0$), and compute the spectra by systematic
expansions around the Ising limit.
We study both spin-half and spin-one
Heisenberg models as well as a range of parameters to cover various
cases proposed for the iron pnictide materials. 
The low-energy spectra have anisotropic spin-wave velocities and are
renormalized with respect to linear spin-wave theory by up to 20 percent,
depending on parameters.
Extreme anisotropy,
consisting of a ferromagnetic $J_{1b}=- |J_F|$, is best distinguished
from a weak anisotropy ($J_{1a}\approx J_{1b}=J_1$, $J_2>J_1/2$ )
by the nature of the spin-waves near the wavevectors
($0,\pi$) or ($\pi,\pi$). The reported spectra for the
pnictide material CaFe$_2$As$_2$ clearly imply
such an extreme anisotropy.

\end{abstract}


\maketitle

The parent phases of iron pnictide superconductors have been found to be metallic
but with antiferromagnetic order at low temperatures.\cite{kamihara,cruz,dong}
There is an ongoing debate between the validity of a strong-coupling picture, 
with local spins interacting via Heisenberg exchange interactions,
and a weak coupling picture where partial nesting of the fermi-surface
leads to a spin-density-wave order.\cite{yin,cao,ma,yildrim,wu,haule,si,yao,xu,mazin,raghu,ran,baskaran,han,rajiv,chen} 
In this paper we will not get into this debate but rather focus on
the systematic calculations of spin-wave spectra for Heisenberg models
on an anisotropic square-lattice with nearest and second neighbor interactions, 
using series expansion methods.\cite{book,advphy} Such studies of spatially
anisotropic interactions on triangular-lattices have proved fruitful in
understanding magnetic properties of several organic and 
inorganic materials.\cite{coldea1,coldea2} This work should similarly be helpful
for understanding materials with an orthorombic square-lattice geometry.\cite{pardini}

Neutron scattering spectra for the pnictides 
show sharp spin-waves.\cite{zhao1}
In the low temperature phase there is orthorombic
distortion and the exchange constants have been found to be substantially
anisotropic. For different materials, and sometimes even for the same
material, different exchange constants have been reported.\cite{zhao,broholm} 
In some cases,
there are reports of extreme anisotropy in the nearest-neighbor exchange.
They are found to be strong and antiferromagnetic along one axis and weak
and ferromagnetic along the other.\cite{zhao} The
origin of the strong spatial anisotropy remains controversial, one theory being that 
it is due to orbital order,\cite{frank,yildrim2,rajiv,chen,ashvin,weiku} 
which may drive the tetragonal to orthorombic transition
in these materials. In this paper, we focus on temperatures much
below the ordering temperature, where in the strong coupling picture
a Heisenberg Hamiltonian should be appropriate.

\begin{figure}
\begin{center}
 \includegraphics[width=0.8\columnwidth,clip,angle=0]{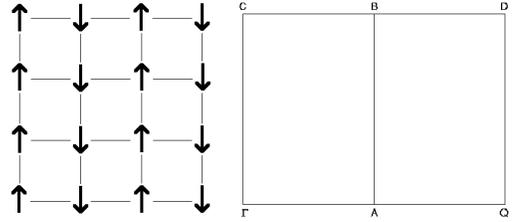}
\caption{\label{fig:Fig1} 
On the left is the ordering pattern of the pnictides. On the
right is the
positive quadrant of
the square-lattice Brillouin Zone showing wavevectors $\Gamma$ ($0,0$),
$A$ ($\pi/2,0$), $Q$ ($\pi,0$), $D$ ($\pi,\pi$), $B$ ($\pi/2,\pi$) and
$C$ ($0,\pi$). With long-range order at ($\pi,0$) 
the distinct energies are contained inside
the region $\Gamma A B C$.
}
\end{center}
\end{figure}

We will consider the Hamiltonian:
\begin{eqnarray}
H =&J_{1a} \sum_i  \vec S_i\cdot \vec S_{i+\hat x}
              + J_{1b}\sum_i  \vec S_i\cdot \vec S_{i+\hat y} \nonumber \\
&+J_2 \sum_{<i,k>} \vec S_i \cdot \vec S_k,\nonumber\\
\end{eqnarray}
The first two terms represent the nearest-neighbor exchange along the
x and y axes respectively. The third term is the second-neighbor exchange
which is taken to be independent of direction. Here we are interested in
parameter ranges that lead to antiferromagnetic order at ($\pi,0$)
as found in the iron pnictides. There are two ranges of parameters
of interest: (i) $J_{1a}$ is the largest energy scale, $J_{1b}$ is
small positive or negative and $J_2$ is of order or smaller than $J_1/2$.
(ii) $J_{1a}$ and $J_{1b}$ are comparable and
$ J_2>J_{1a}/2, J_{1b}/2$. The latter case is highly frustrated and
colinear ($\pi, 0$) order is stabilized by quantum fluctuations.
In the former case, the system is unfrustrated or weakly
frustrated and ($\pi,0$) order
minimizes all or nearly all the interactions. 

The case of $J_{1a}=J_{1b}$ was discussed in an earlier study.\cite{uhrig}
That case is conceptually more subtle as the classical ground state
in the ($\pi,0$) phase 
is highly degenerate. Spins on the two
sublattices of the square-lattice are free to rotate with
respect to each other. The colinear order is selected by
quantum fluctuations through 
an order by disorder mechanism.\cite{shender,chandra,capriotti} This also has important
consequences for the spin-wave spectra. The linear spin-wave
spectra has spurious gapless modes in addition to those
required by Goldstone's theorem. These become gapped
upon proper inclusion of quantum fluctuations.\cite{uhrig,singh} Once
$J_{1a}$ is not equal to $J_{1b}$, the classical
ground state becomes unique becoming antiferromagnetic
along the direction of larger exchange, and linear spin-wave
theory should give the qualitatively correct spectra.

The linear spin-wave dispersion for the model is given by\cite{rajiv}
\begin{equation}
\omega_k= 4 S J_2 \sqrt{(A_k^2-B_k^2)}
\end{equation}
with
\begin{equation}
A_k=1+\alpha-\beta+\beta \cos{k_y},
\end{equation}
and,
\begin{equation}
B_k=\cos(k_x) (\cos(k_y)+\alpha).
\end{equation}
Here, $\alpha=J_{1a}/(2 J_2)$, and $\beta= J_{1b}/(2 J_2)$. 
The spectral weights associated with the spin-waves is given by the expression
\begin{equation}
S_k\propto \sqrt{{(A_k-B_k)\over(A_k+B_k)}}
\end{equation}
These lead to spin-wave velocity along $x$ of 
$$v_x=2S (2J_2+J_{1a})$$
and along $y$ of 
$$v_y=2S \sqrt{(2J_2-J_{1b})(2J_2+J_{1a})}.$$

For the numerical calculations, it is convenient to set $J_{1a}=1$. 
The actual energy scale for the material can be deduced by comparing
with experiments. Motivated by the
experimentally reported parameters,\cite{zhao,broholm} 
we will study five different parameter
sets: (i) $J_{1b}=-0.2$, $J_{2}=0.4$,
(ii) $J_{1b}=0$, $J_{2}=0.4$,
(iii) $J_{1b}=0.2$, $J_{2}=0.4$,
(iv) $J_{1b}=0.2$, $J_{2}=0.9$,
(v) $J_{1b}=0.8$, $J_{2}=1.4$. In all cases, will calculate spectra for both
spin-half and spin-one models to see if the shape of the spectra has any significant
spin dependence. 

For all these parameters, we develop Ising
series expansions for the spin-wave dispersion and their spectral weights.
The series are computed to 8-th order and involve a set of 280474 distinct
clusters.
These are analyzed throughout the zone using series extrapolation methods.
These extrapolation methods converge extremely well if one is not too close to
($0,0$) or the ordering wavevector ($\pi,0$). The dispersion must go to
zero near these, with a linear in $q$ behavior although with anisotropic
spin-wave velocities. We have used the method of Singh and Gelfand\cite{sg} 
to calculate the spin-wave velocities. Very near these wavevectors the linear
dispersion is assumed with the calculated anisotropic spin-wave velocities
to obtain the spectra.
The spectral weights are calculated by the methods discussed by
Zheng et al.\cite{zheng}
The spectral weights vanish
near ($0,0$) but diverge as $1/q$ near the ordering wavevector. We
will not focus much on the region very close to this divergence. 
Away from that point, simple Pade approximants (or just addition of
terms in the series) converges very well. We will
see that what distinguishes the different models, after an overall
energy scale has been scaled out of the problem, is the nature
of the high-energy short-wavelength spin-waves and that is our
primary focus here.

The spin-wave velocities along x and y for the different parameter
ranges calculated from the series expansions are shown in Table I for 
spin-half and Table II for spin-one.
The colinear ordering pattern and
the square-lattice Brillouin zone with some q-vectors used for
defining the contours
along which spectra will be shown are depicted in Fig~1. 

\begin{figure}
\begin{center}
 \includegraphics[width=0.8\columnwidth,clip,angle=0]{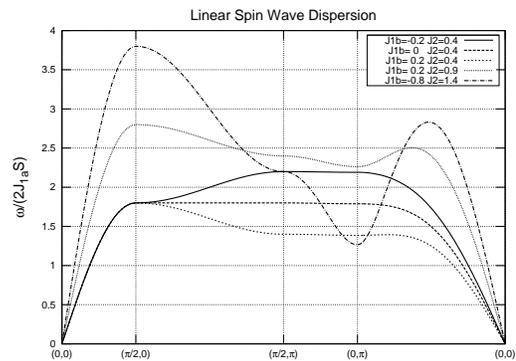}
\caption{\label{fig:Fig2} Linear spin-wave spectra for the models
along selected contours in the Brillouin Zone.}
\end{center}
\end{figure}

\begin{figure}
\begin{center}
 \includegraphics[width=0.8\columnwidth,clip,angle=0]{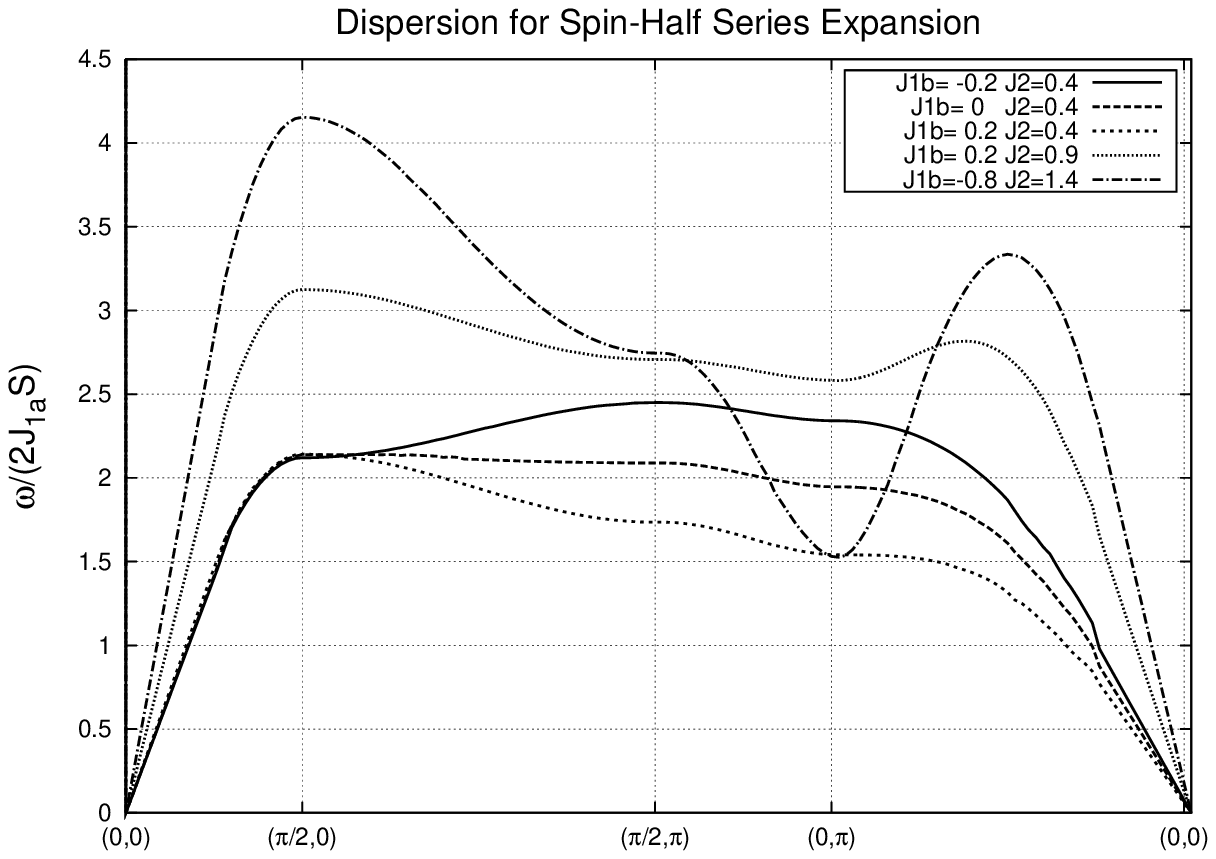}
\caption{\label{fig:Fig3} Spin-wave spectra calculated by series expansions for the spin-half 
models along selected contours in the Brillouin Zone.}
\end{center}
\end{figure}

\begin{figure}
\begin{center}
 \includegraphics[width=0.8\columnwidth,clip,angle=0]{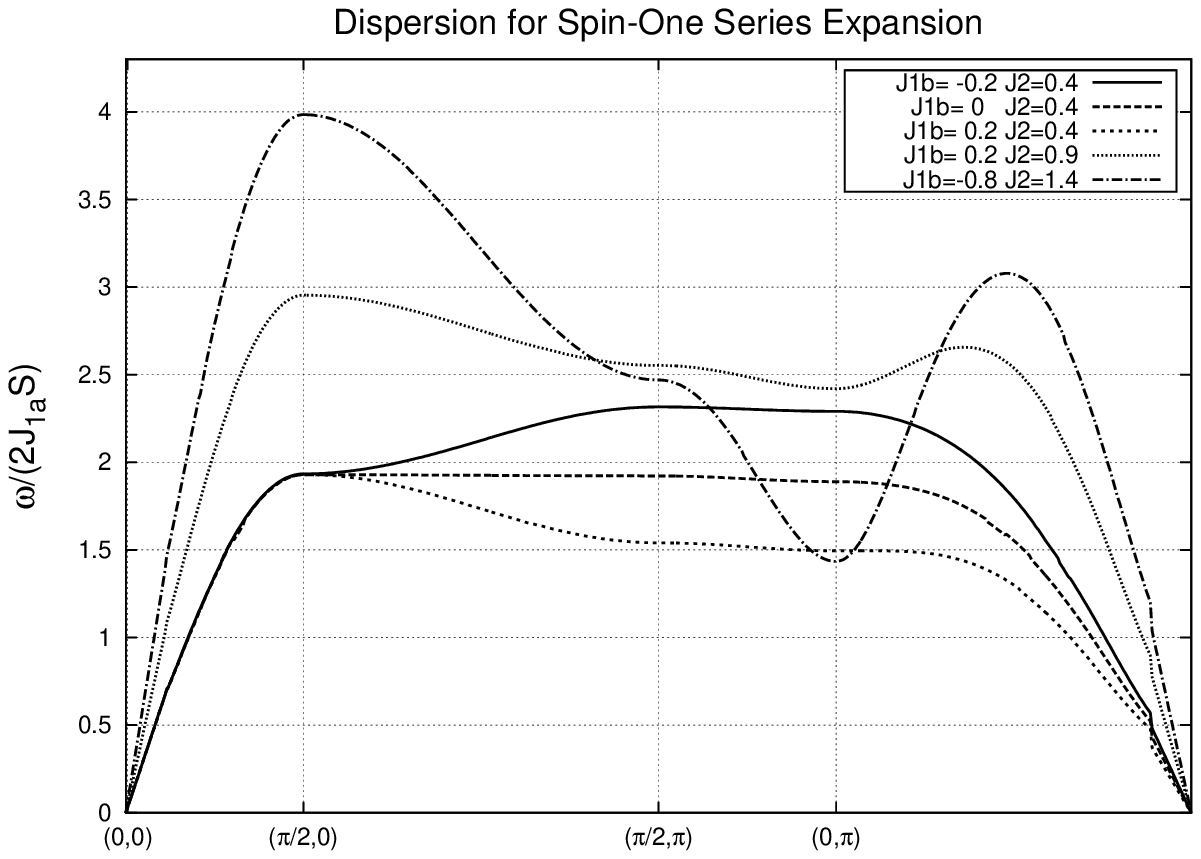}
\caption{\label{fig:Fig4} Spin-wave spectra calculated by series expansions for the spin-one 
models along selected contours in the Brillouin Zone.}
\end{center}
\end{figure}

\begin{figure}
\begin{center}
 \includegraphics[width=0.8\columnwidth,clip,angle=0]{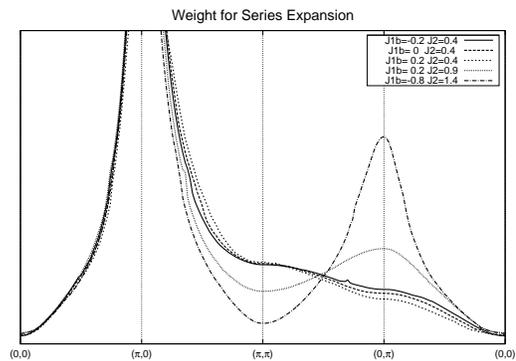}
\caption{\label{fig:Fig5} Spectral weights associated with the spin-waves, in arbitrary units,
along a special contour in the momentum space
for the spin-half Heisenberg models, as calculated by 
Ising series expansions.
}
\end{center}
\end{figure}

In Fig~2, Fig.~3, and Fig.~4, we show the calculated spectra along a selected
contour in the Brillouin zone for the spin-half model, spin-one model,
and linear spin-wave theory respectively. 
The uncertainties in the series calculations, over most of the
Brillouin zone, are of order one percent.
Note that within linear spin-wave theory spin-wave dispersion would
be independent of spin once an overall energy scale has been taken out.
There is a clear overall similiraity
between the spectra, showing that spin value does not significantly alter
the shape of the spectra. 
Also, having $J_{1a}$ not equal to $J_{1b}$ clearly improves the validity of
spin-wave theory.\cite{uhrig}
The primary correction to
linear spin-wave theory is an upward renormalization of the spectra,
which is up to $20\%$ for the spin-half case and less than $10\%$
for the spin-one case. Even at low energies these renormalizations
are found to be anisotropic.
The renormalization of spin-wave energy is especially non-uniform
near the antiferromagnetic zone-boundary. Most notably, flat regions of
the linear spin-wave spectra acquire some dispersion on inclusion
of quantum fluctuations. As expected, these structures are more pronounced for
spin-half than for spin-one case.
This is not dissimilar to
the nearest-neighbor square-lattice case, where also the zone-boundary
dispersion acquires a structure that is absent in linear spin-wave 
theory.\cite{sg,zheng,sandvik}

\begin{figure}
\begin{center}
 \includegraphics[width=0.8\columnwidth,clip,angle=0]{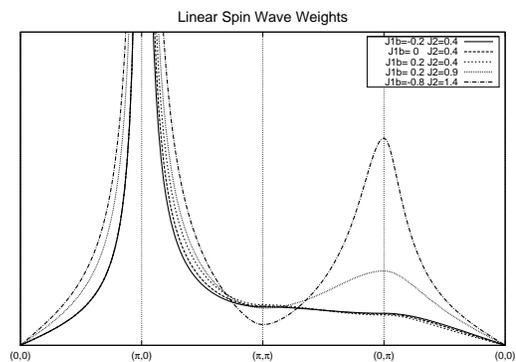}
\caption{\label{fig:Fig6} Spectral weights associated with the spin-waves, in arbitrary units,
along a special contour in the momentum space obtained from Linear Spin Wave theory}
\end{center}
\end{figure}

The spectral-weights associated with the spin-waves for the
spin-half models and for linear spin-wave theory are shown in
Fig.~5 and Fig.~6 respectively. One finds that along cetrain directions,
and especially at long-wavelengths the different models
are indistinguishable. The major differences between different
parametrs sets arise when one
considers the weights at short wavelengths or high energies. 
In the extreme anisotropy
case, when the spin-wave is a maximum at ($0,\pi$), there is
only a small scattering intensity around that wavevector.
In the weak anisotropy limit, when there is low excitation energies
at these wavevectors, there is also enhanced intensity
at these wavevectors. 

The density of states for the spin-half models are shown in Fig.~7.
The key distinguishing feature is that
the weakly frustrated models have sharp peaks close to highest
energies. This is also evident from the spectra, where there are
flat regions in the dispersion curve.

\begin{figure}
\begin{center}
\includegraphics[width=0.8\columnwidth,clip,angle=270]{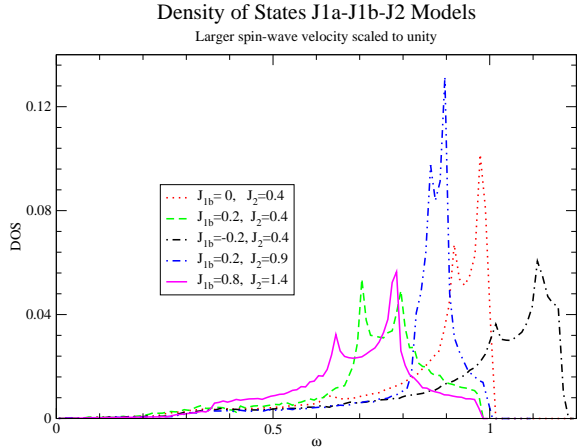}
\caption{\label{fig:Fig7} Density of states for the different 
spin-half models as calculated by series expansions.
}
\end{center}
\end{figure}

We now discuss the relevance of these calculations to the observed
spectra in the iron pnictide materials. We first note that the observation
of sharp spin-waves throughout the Brillouin zone would be strongly
supportive of a local moment picture. Zhao et al\cite{zhao} have argued that
this is indeed the case and that there is an absence of a Stoner continuum
in these materials, which should have been present if 
an itinerant picture for the magnetism was more appropriate. 
This suggests that magnetism and metallic behavior can be treated
separately.
This controversial issue\cite{tom} 
is clearly beyond the scope of the present work.
We will instead restrict ourselves to discussing the
spin-wave dispersion and spectral
intensities expected from the Heisenberg models 
in different parameter regimes, so that they
can guide future experiments.
As discussed earlier, spin does not play a big role in these models except
to set the overall energy scale in terms of $J$. However, since $J_{1a}$
is an adjustable parameter, it can always be rescaled to match the
experimental data. Hence, all
comparisons below are done in terms of the spin-half models.

\begin{figure}
\begin{center}
 \includegraphics[width=8cm]{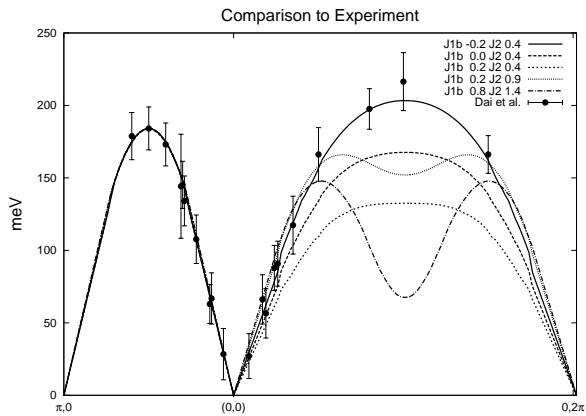}
\caption{\label{fig:Fig8} A comparison of the measured spectra in CaFe$_2$As$_2$,
with the different models.
}
\end{center}
\end{figure}

Fig.~8, shows a comparison of the calculated dispersion for the
different parameters with the experimental measurements in CaFe$_2$As$_2$.
In all cases, the exchange constant $J_{1a}$ is adjusted to match
with the low energy spectra near the zone center. The estimated
values for $J_{1a}$ are $86.9$, $86.1$, $86.0$, $58.9$ and $44.3$ meV
in the cases (i) through (v) respectively. Note that theoretical
error bars are of order one percent.
It is clear that 
all parameters are equally good for describing
the zone-center spectra along both x and y axes.
The difference really arises when one studies the
zone-boundary excitations. In particular the spectra near ($0,\pi$)
can only be explained by the parameters $J_{1b}/J_{1a}=-0.2$,
$J_2/J_{1b}=0.4$. Even if we make $J_{1b}$ zero or slightly positive,
we can no longer describe the high energy spectra. The weakly
anisotropic models have sharp dips near ($0,\pi$) and hence have no
chance of describing the observed spectra.
It would be useful to systematically look for the high energy spin-wave
spectra in different family of iron pnictide materials to see how 
universal the high energy spectra is. 

To further guide neutron scattering studies in this direction, we
create two dimensional scattering intensity profiles in the Brillouin zone
at different frequencies. Our results do not include any form-factor effects 
and unlike experiments have no noise.
We use an artificial Gaussian broadening in $\omega$ to
mimic finite experimental resolution. Thus we take
$$S(q,\omega)=S(q)\exp{-{(\omega-\omega_q)^2\over \Delta^2}}.$$
with a suitably chosen $\Delta$, which we take to be independent of $q$.
Here $S(q)$ and $\omega_q$ are the spectral intensity and
spin-wave frequencies calculated by series expansions. We focus
on cases (i) and (v), which correspond to most anisotropic exchanges and
least frustrated model
and least anisotropic exchanges and most frustrated model respectively.

In Fig.~9 , the intensities are plotted over the full Brillouin Zone
for several different frequencies for the models (i) and (v).
The evolution from a single bright spot at the zone center at low energies, 
due to finite resolution, to
an ellipse with a hole in the middle at intermediate energies 
is a standard feature
of this type of ($\pi,0$) order. This feature is similar for all parameter sets.
There are clear differences, however, even at low frequencies, which should be resolvable with high accuracy data.
The plots on the left are more elliptical and those on the right
are more circular.
As one moves to high energies and excitations
move far from the zone center, details of the local Hamiltonian
become clearly visible. The two cases shown have vastly different
spectra. It should be noted that relative to the zone center, the
intensity at higher energies is significantly diminished.
At the highest energy shown, excitations are present only in the
plots on the left. The plots on the right just show weak vestiges of
lower energy excitations due to the assumed finite resolution.

\begin{figure}
\begin{center}
 \includegraphics[width=8cm]{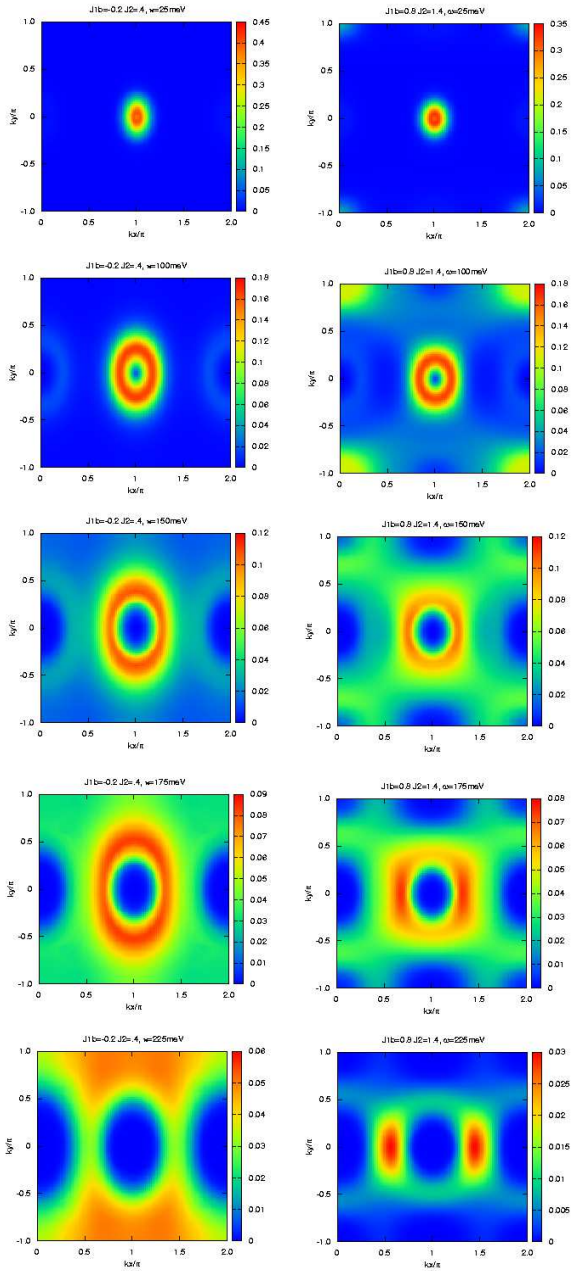}
\caption{\label{fig:Fig9} Scattering intensities in the full Brillouin zone
centered at the ordering wavevector, for (from top to down) 
$\omega=$ 25 meV, 100 meV, 150 meV,
175 meV and 225 meV  for the strongly anisotropic, 
weakly frustrated model
on the left and weakly anisotropic, strongly frustrated model on the right, }
\end{center}
\end{figure}

\begin{figure}
\begin{center}
 \includegraphics[width=8cm]{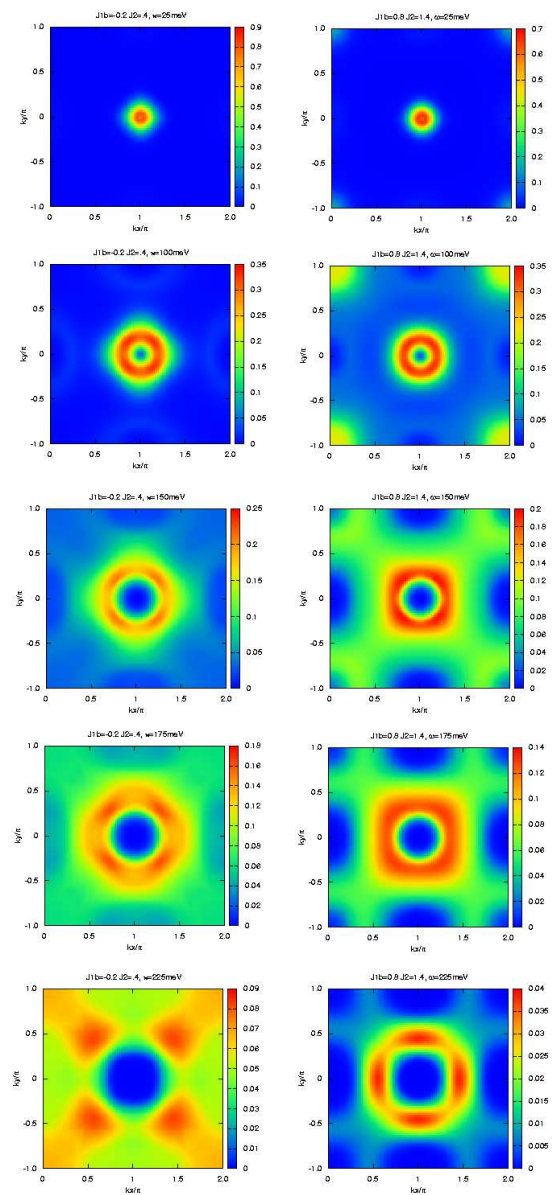}
\caption{\label{fig:Fig10} Scattering intensities as in Fig.~9 in
a substantially twinned sample, which leads to restoration
of tetragonal symmetry}
\end{center}
\end{figure}

In Fig.~10 intensity plots are made upon averaging the spectra
at ($q_x,q_y$) and ($q_y,q_x$), as would be expected in a heavily
twinned sample. It is evident that major distinctions between
the two models remains evident despite the restoration of the 
$90$ degree rotational symmetry.
So, while the
detwinning of the materials may be important to get complete information,
spectra from a twinned sample can also distinguish a weakly anisotropic
model from a strongly anisotropic one.

In conclusion, in this paper we have used series expansion methods to
calculate the spin-wave spectra and spectral weights for orthorombic
square-lattice Heisenberg models. We find that the linear spin-wave theory is
qualitatively valid for weak and strong frustration. This case is
different from a system with tetragonal symmetry where linear spin-wave theory
was found to be qualitatively incorrect. In general, the renormalization
of the spin-wave energies throughout the zone is of order or less
than 20 percent. The high energy spin-waves and their dispersion provide
a particularly sensitive way
to narrow down parameter ranges and determine the extent
of spatial anisotropy in the exchange constants of different orthorombic
materials. 

The spectra of the iron pnictide materials imply a
strongly anisotropic system, where nearest neighbor exchanges are
strong and antiferromagnetic in one direction and weak and 
ferromagnetic in the other. 
While this study has ignored the metallic nature of the pnictides
and focused entirely on a local moment description,
the conclusion of strong spatial anisotropy is likely to
have much broader validity.
The implications of this anisotropy in
other properties of the system deserve further attention.



\begin{acknowledgements}
We would like to thank G. Uhrig, O Sushkov, S. Savrasov and W. Pickett
for useful discussions.
\end{acknowledgements}


\begin{table}[h]
\caption{[m/n] Pade and overall estimates (Est.) of spin wave velocities ($v_x,v_y$)
for spin-half models for different $J_{1b}$ and $J_2$ values with $J_{1a}=1.0$}
\begin{tabular}{rrrrrrrr}
\hline\hline
J$_{1b}$, J$_2$ & v& [4/4] & [5/3] & [3/5] & [4/3] & [3/4] & Est.\\
-0.2, 0.4  & v$_x$  & 2.1029 & 2.1043 & 2.1039 & 2.1013 & 2.1012 & 2.10 \\
-0.2, 0.4  & v$_y$  & 1.4162 & 1.4123 & 1.4114 & 1.4056 & 1.4016 & 1.41 \\
 0.0, 0.4  & v$_x$  & 2.1169 & 2.1244 & 1.9803 & 2.1360 & 2.1268 & 2.12 \\
 0.0, 0.4  & v$_y$  & 1.2590 & 1.3712 & 1.1754 & 1.2622 & 1.2619 & 1.26 \\
 0.2, 0.4  & v$_x$  & 2.1120 & 2.3405 & 2.1718 & 2.3117 & 2.1726 & 2.19 \\
 0.2, 0.4  & v$_y$  & 1.0749 & 1.1355 & 1.1092 & 1.1645 & 1.1200 & 1.11 \\
 0.2, 0.9  & v$_x$  & 3.1269 & 3.1532 & 3.1444 & 3.1434 & 3.1410 & 3.14 \\
 0.2, 0.9  & v$_y$  & 2.3616 & 2.3804 & 2.3771 & 2.3816 & 2.3796 & 2.38 \\
 0.8, 1.4  & v$_x$  & 4.1823 & 4.3840 & 4.2505 & 4.2848 & 4.2253 & 4.26 \\
 0.8, 1.4  & v$_y$  & 3.2355 & 3.3711 & 3.2967 & 3.4350 & 3.3278 & 3.33 \\
\hline\hline
\end{tabular}
\end{table}

\begin{table}[h]
\caption{[m/n] Pade and overall estimates (Est.) of spin wave velocities ($v_x,v_y$)
for spin-one models for different $J_{1b}$ and $J_2$ values with $J_{1a}=1.0$}
\begin{tabular}{rrrrrrrr}
\hline\hline
J$_{1b}$, J$_2$ & v& [4/4] & [5/3] & [3/5] & [4/3] & [3/4] & Est.\\
-0.2, 0.4  & v$_x$/4  & 0.9669 & 0.9669 & 0.9692 & 0.9684 & 0.9676 & 0.968 \\
-0.2, 0.4  & v$_y$/4  & 0.6938 & 0.6986 & 0.6999 & 0.6959 & 0.6950 & 0.697 \\
 0.0, 0.4  & v$_x$/4  & 0.9650 & 1.0128 & 0.9712 & 0.9698 & 0.9692 & 0.970 \\
 0.0, 0.4  & v$_y$/4  & 0.6202 & 0.6567 & 0.6256 & 0.6236 & 0.6227 & 0.624 \\
 0.2, 0.4  & v$_x$/4  & 0.9673 & 0.9696 & 0.9692 & 0.9701 & 0.9696 & 0.969 \\
 0.2, 0.4  & v$_y$/4  & 0.5370 & 0.5398 & 0.5389 & 0.5363 & 0.5398 & 0.539 \\
 0.2, 0.9  & v$_x$/4  & 1.4797 & 1.4828 & 1.4819 & 1.4834 & 1.4819 & 1.482 \\
 0.2, 0.9  & v$_y$/4  & 1.1195 & 1.1226 & 1.1217 & 1.1235 & 1.1217 & 1.122 \\
 0.8, 1.4  & v$_x$/4  & 2.0100 & 1.9999 & 1.9967 & 1.9932 & 1.9900 & 1.999 \\
 0.8, 1.4  & v$_y$/4  & 1.5016 & 1.5026 & 1.5016 & 1.5041 & 1.5015 & 1.502 \\
\hline\hline
\end{tabular}
\end{table}
\end{document}